\documentclass[a4paper]{jpconf}
\usepackage{graphicx}
\usepackage[OT2,T1]{fontenc}
\usepackage[utf8]{inputenc}
\usepackage[russian,english]{babel}

\def\lapp{\ifmmode\stackrel{<}{_{\sim}}\else$\stackrel{<}{_{\sim}}$\fi}
\def\gapp{\ifmmode\stackrel{>}{_{\sim}}\else$\stackrel{>}{_{\sim}}$\fi}
\def\degr{\ifmmode^{\circ}\else$^{\circ}$\fi}

\begin{document}
\title{50 Years of Pulsars!}

\author{R. N. Manchester}

\address{CSIRO Astronomy and Space Science, PO Box 76, Epping NSW 1710, Australia}

\ead{dick.manchester@csiro}

\begin{abstract}
  A brief, personal, and very incomplete account of 50 years of pulsar
  astronomy presented at the Conference Dinner for 
  ``Physics of Neutron Stars -- 2017 --  50 Years After'', held in
  Saint Petersburg, July 2017.
\end{abstract}

\section{Introduction and the discovery of pulsars }
First, I would like to thank George Pavlov for the invitation to
attend this meeting and to present this talk at the Conference
Dinner. I have to say, I was a little surprised to be asked to give
the after-dinner talk as I am not well-known as a raconteur. I could only
conclude that it was because I am old!

It is true that I have been involved with pulsars one way or another
since their discovery, or at least the official announcement of it by
Hewish et al. in the February 24, 1968, edition of Nature. I started
work at the CSIRO Parkes Observatory just 12 days before the Nature
paper was published and soon became involved in pulsar research.

\begin{figure}[h]
\includegraphics[width=130mm]{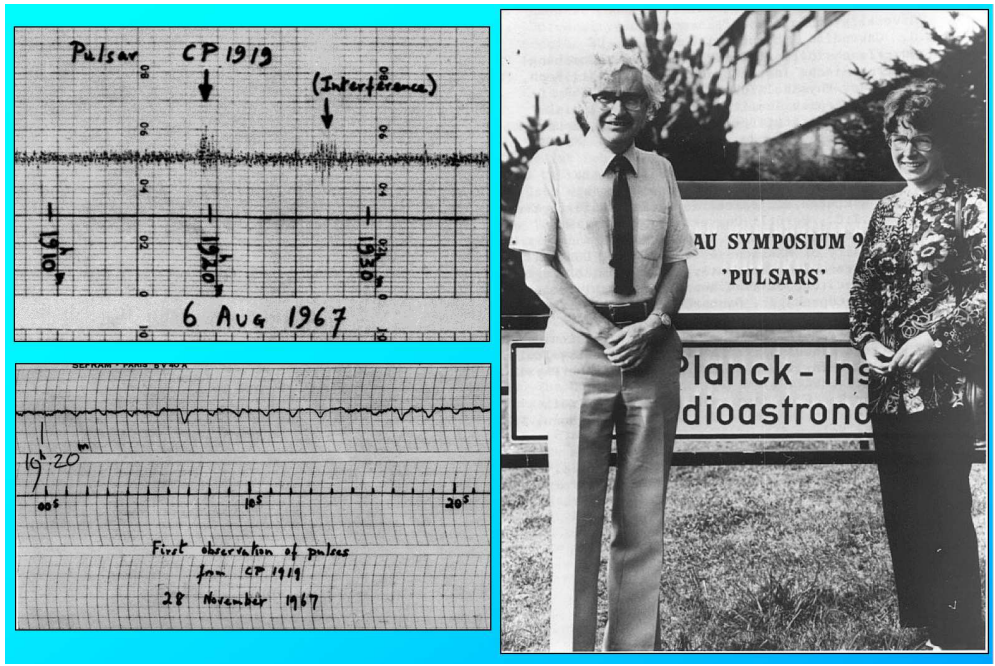}
\caption{Left: Chart recordings from the Four-acre array at Cambridge
  showing the first detected signals from CP 1919 and the first
  recording to show that the emission was pulsed. (Reprinted with
  permission from Reviews of Modern Physics \cite{hew75a}, copyright (1975) by the American
  Physical Society.) Right: Jocelyn Bell and Antony Hewish at IAU
  Symposium 95, Bonn, August 1980. }\label{fg:disc}
\end{figure}

But first back to the real beginning. In mid-1967, Jocelyn Bell was a
graduate student at Cambridge University helping to build the
``Four-acre array'', an array of dipoles tuned to 81.5 MHz that Antony
Hewish had designed to investigate interplanetary scintillation of
radio sources with the aim of identifying compact sources likely to be
quasars. With the completion of the array, Jocelyn was given the job
of examining the chart recordings to find the rapidly fluctuating
signals from compact sources that scintillated as a result of
propagation through the solar wind. She did find many of these, but
also began to notice some signals that were subtly different from both
the scintillating sources and radio-frequency interference. The first
recorded detection of one of these ``scruff'' sources was on August 6,
1967, as shown in Figure~\ref{fg:disc}. The break-through and the
point that marks the real discovery of pulsars was when she realised
that this source returned at the same sidereal time most days. This
meant that it had to be outside the solar system and unrelated to
terrestrial activity. A few months later, on November 28, she and
Antony Hewish obtained the first high-speed recording to show that the
scruff was in fact a train of pulses with a spacing of about 1.3
seconds.

Follow-up observations with the Four-acre array and other instruments
at Cambridge refined the periodicity of the pulsar and showed that the
signals were dispersed by their passage through the interstellar
medium. A paper reporting the discovery and follow-up observations of
the pulsar, then named CP 1919, and an interpretation that favoured
radial pulsations of either a white dwarf or a neutron star was
submitted to Nature by Hewish, Bell, et al. on February 9, 1968, and
published on February 24 \cite{hbp+68}.

In 1974 Antony Hewish was awarded the Nobel Prize in Physics for the
discovery of pulsars. 

\section{Pulsar observations at Parkes}
\begin{figure}[h]
\includegraphics[width=100mm]{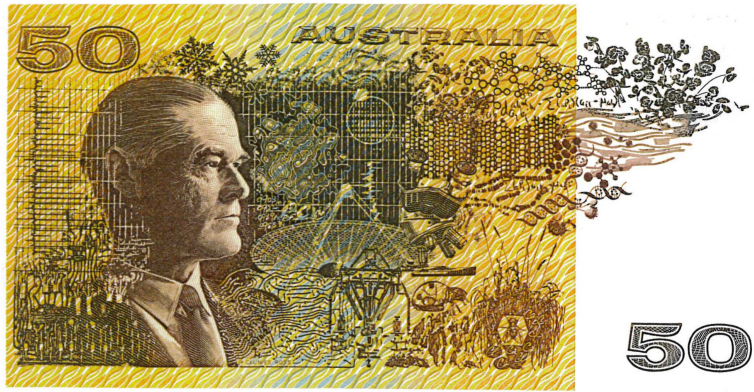}
\caption{The first Australian \$50 note, featuring the Parkes radio
  telescope, a portrait of Sir Ian Clunies Ross (the first Chairman of
  CSIRO) and various images relating to Australian radio astronomy and
  biological sciences. On the left of the note is an image of the
  first pulsar recording made at Parkes, of CP 1919 and recorded on
  March 8, 1968.}\label{fg:50AUD}
\end{figure}

Just two weeks after the publication of the Nature discovery paper a
team from the CSIRO and the University of Sydney installed a
remarkable set of five coaxial feeds with corresponding receivers at
the focus of the Parkes 64-m radio telescope. On the morning of March
8, I was in the control room as the telescope slewed around to the
position of CP 1919 and saw the wonderful burst of pulses come through
on the chart recorder seconds later. By good fortune, this first
observation at 150 MHz just caught a large scintillation maximum. The pulsar
was not seen again with such high signal-to-noise for the rest of the
session. This image was recorded for posterity on the first Australian
\$50 note (Figure~\ref{fg:50AUD}) and also published with what are
still some of the best measurements of the individual-pulse spectra of
any pulsar \cite{rcg+68}.

\begin{figure}[h]
\includegraphics[width=100mm]{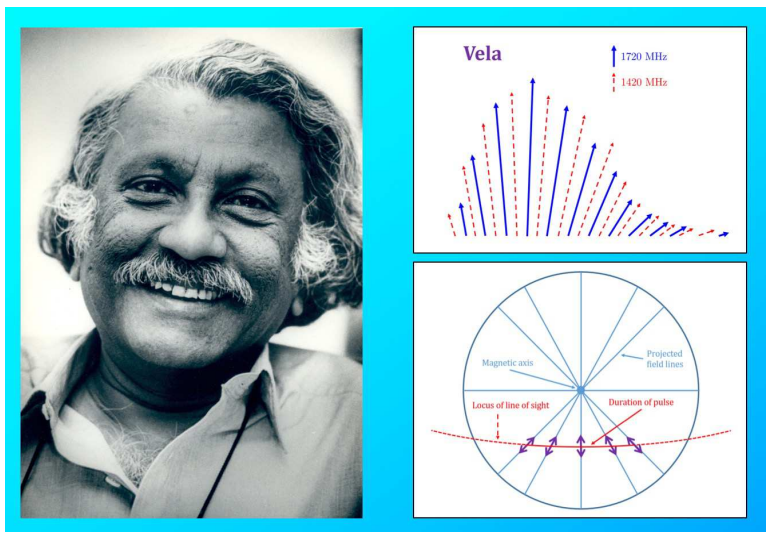}
\caption{Left: Radhakrishnan in the year 1996 (image credit: Library,
  Raman Research Institute). Right: plots illustrating the
  frequency-independence of the position-angle swing across the Vela
  pulsar mean pulse profile and the interpretation of that in terms of
  the magnetic-pole or rotating-vector model (after Radhakrishnan \&
  Cooke \cite{rc69a}). In this model, the radiation is emitted
  tangentially to field lines and is linearly polarised in the plane of
  field-line curvature as indicated by the double-headed arrows in the
  lower plot. (Image credit: Peter Shternin)}\label{fg:rvm}
\end{figure}

Later on that year, having learnt something about measuring
polarisation with the 18-cm receiver while working with Brian Robinson
and Miller Goss on observations of OH masers, I was asked to assist
Radhakrishnan investigate the polarisation of the Vela pulsar. These
observations led to rotating-vector model (RVM) (or magnetic-pole
model as Rad preferred to call it) for pulsar polarisation
(Figure~\ref{fg:rvm}) \cite{rc69a}. In follow-up observations in
mid-March, 1969, we again pointed the telescope to the Vela pulsar,
recording the data with a signal-averaging system that was set up to
fold the data at the expected topocentric period of the pulsar. To our
surprise, rather than remaining a fixed phase on the oscilloscope
display, the pulse was noticeably drifting to the left, indicating an
error in the predicted period. By the evening, after a while trying to understand what
was wrong, Rad said that he was going to bed and left me to
sort out the problem. I again checked all the instrumentation, looked at
some other pulsars and eventually concluded that it must be the
pulsar that had changed. I left a note for Rad and went to bed
myself.

Observations over the next few days confirmed that it was a
real decrease in the pulsar period by about 3 parts in a million - the
first detection of a pulsar glitch! Rad contacted Paul Reichley and
George Downs who we knew were observing the pulsar using the Goldstone
antenna at the Jet Propulsion Laboratory, Caltech, and found that they
too had observed the glitch. Back-to-back papers reporting the
discovery were published in Nature on April 19, 1969 \cite{rm69,rd69}.

\section{Searching for pulsars}
Searches for pulsars are fundamentally important. Not only does the
discovery of previously unknown pulsars increase the sample for all
manner of statistical studies of pulsar properties and evolution, but
almost every significant pulsar search has uncovered some unexpected
and interesting pulsar or class of pulsars. Examples are binary
pulsars \cite{ht75a}, pulsars in globular clusters \cite{lbm+87},
RRATs \cite{mll+06}, pulsars with planets \cite{wf92}, pulsars with
main-sequence binary companions \cite{jml+92}, etc., etc..

\begin{figure}[h]
\includegraphics[width=100mm]{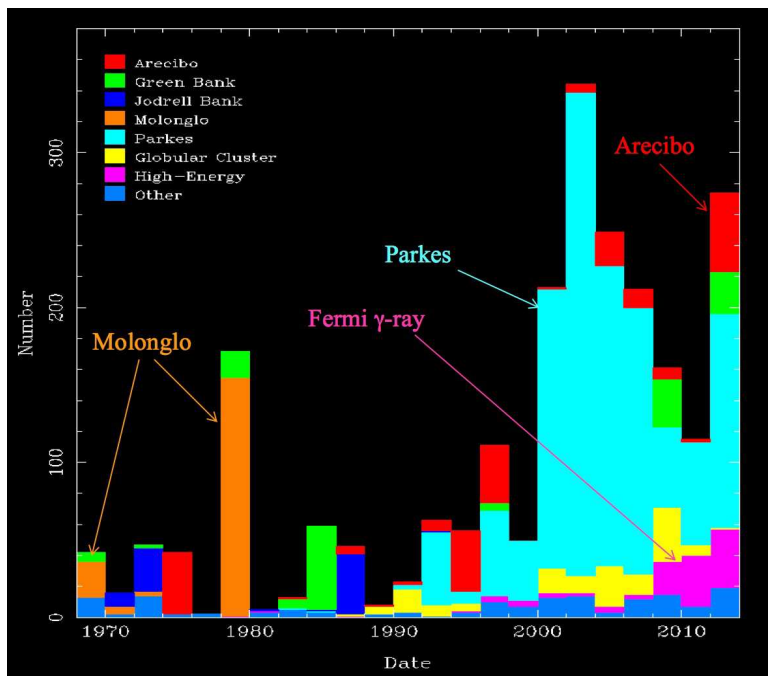}
\caption{Rate of pulsar discovery, sorted by observatory or
  class. Each bar represents the number of pulsars discovered in the
  corresponding 2-year interval. Data from the ATNF Pulsar Catalogue
  \cite{mhth05}}\label{fg:search}
\end{figure}

More than 2600 pulsars are now known. Figure~\ref{fg:search} shows the
rate of pulsar discovery since 1968 by observatory (or in a couple of
cases, pulsar class). Small searches (including the original Cambridge
search) are grouped under ``Other''. This figure highlights the major
contributions made by the Molonglo radio telescope in the first decade
-- twice in this time, more than half of the known pulsars were discovered
at Molonglo -- and the Parkes radio telescope. In the late 2000's,
Parkes had found more than twice as many pulsars as the rest of the
world's telescopes put together. Even now, the Parkes share is more
than 55\% of the total. The Parkes Multibeam Pulsar Survey
\cite{mlc+01} by itself has more than 830 pulsars to its credit,
including recent discoveries from reanalyses of the dataset, e.g.,
\cite{kek+13}. In recent years, the {\it Fermi} Gamma-ray Space
Telescope has been very successful in uncovering previously unknown
pulsars, especially millisecond pulsars (MSPs) in so-called
``redback'' and ``black widow'' binary systems where the pulsar radio
emission is often obscured or eclipsed by plasma streams from the
companion star. Most of these systems have been found through radio
searches of unidentified gamma-ray sources with properties known to
be characteristic of pulsars, e.g., \cite{ckr+15}.

\begin{figure}[h]
\includegraphics[width=90mm]{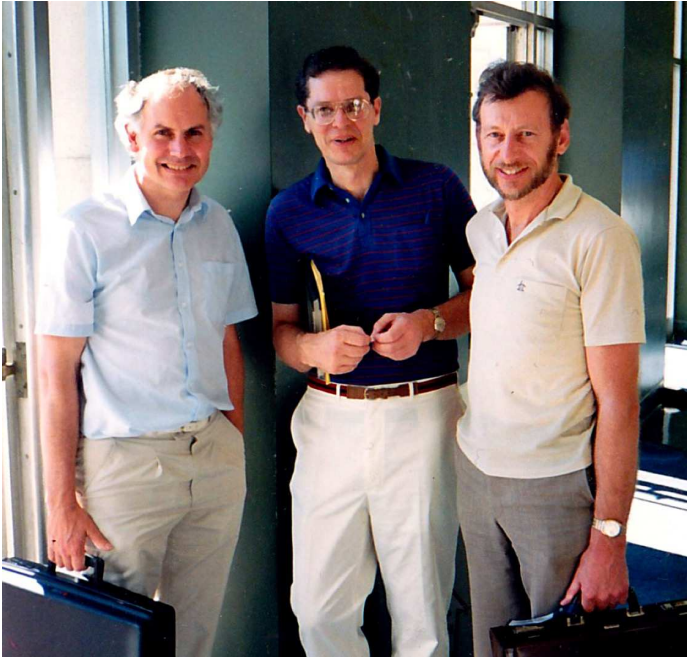}
\caption{Andrew Lyne, Joe Taylor and Dick Manchester at IAU Colloquium
  177 ``Pulsar Astronomy -- 2000 and beyond'', Bonn, August,
  1999.}\label{fg:lmt}
\end{figure}

People often ask me why Parkes has been so successful in pulsar
discovery. There are three main reasons:
\begin{itemize}
\item The centre of our Galaxy passes almost overhead at Parkes and
  the southern part of the Galactic plane is far richer than that in
  the north. 
\item At ATNF we have had and continue to have a very skilled group of
  engineers and excellent co-operation with the scientific staff. This
  led, for example, to the development of the Parkes 13-beam receiver
  system, by far the most successful pulsar-finding machine ever.
\item We had very experienced teams of scientists working on the
  pulsar survey projects at Parkes (see, e.g.,
  Figure~\ref{fg:lmt}). This led to the development of efficient
  signal-processing systems (both hardware and software) for pulsar
  detection and confirmation.
\end{itemize}

\section{Key discoveries}
The discovery of the first binary pulsar, PSR B1913+16, at Arecibo in
1974 by Russell Hulse and Joe Taylor \cite{ht75a} is the prime example
of an unexpected and exciting find in pulsar searches. Observations of
this system over the next few years provided the first accurate
determinations of neutron-star masses, the first observational
evidence for gravitational radiation and confirmation that Einstein's
general theory of relativity gives an accurate description of motions
in strong gravitational fields \cite{tfm79,wnt10}. These important
results led to the award of the 1993 Nobel Prize in Physics to Taylor
and Hulse. As an aside, I note that I shared an office with Joe at the
University of Massachusetts from 1971 to 1974, but unfortunately was
not involved in the Arecibo search!

\begin{figure}[h]
\includegraphics[width=120mm]{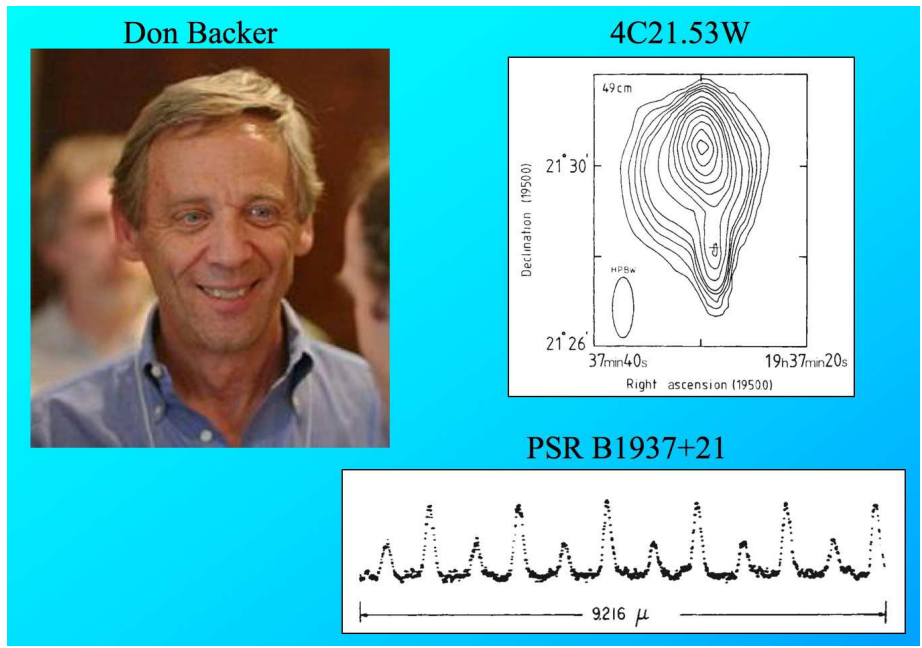}
\caption{The discovery of the first MSP: Don Backer who led the team
  (image credit: Bob Rood), a 600-MHz image of 4C21.53W made with the
  Westerbork Synthesis Radio Telescope, and a signal-averaged
  recording from Arecibo showing the 1.56~ms periodicity. The cross
  marks the position of the compact source in the Westerbork image.
  (Reprinted by permission from Macmillan Publishers Ltd: Nature,
  \cite{bkh+82}, copyright 1982.)}\label{fg:msp1}
\end{figure}

Probably the next most important pulsar discovery was that of the
first MSP at Arecibo by Don Backer and his colleagues
\cite{bkh+82}. This was a little different to the other examples of
important discoveries in that it was not found in a large-scale
search, but rather in a directed investigation of a highly unusual radio
source, 4C21.53W. This source was known to have a compact component
(it scintillated) and it is steep-spectrum and polarised, all
properties consistent with it being a pulsar. Yet efforts to detect a
pulsar in this direction (including one at Parkes by Nichi D'Amico and
RNM) had been unsuccessful. This changed dramatically when Backer et
al. employed a sub-millisecond sampling system that revealed the
1.558~ms periodicity. Figure~\ref{fg:msp1} shows a photograph of Don
Backer, probably taken in the early 2000's, and images from the
discovery paper. Papers suggesting that the rapid spin of the pulsar
originated in an earlier phase of accretion from a companion star
(since disappeared) were quickly published \cite{rs82,acrs82}. The
``recycling'' idea had previously been invoked to account for the
short period of PSR B1913+16 \cite{sb76}.

Finally, I must mention the Double Pulsar, PSR J0737$-$3039A/B,
discovered at Parkes in 2003 \cite{bdp+03,lbk+04}. This system is
unique in a number of ways. Firstly it is a double-neutron-star (DNS) system
like PSR B1913+16, but with the very short orbital period of
2.4~h. This makes it even more ``relativistic'' than PSR B1913+16,
with a predicted (and observed) periastron advance of
$16.9\degr$~yr$^{-1}$, four times that of PSR B1913+16. Secondly, it
was and remains the only DNS system where both stars have been
observed as pulsars. The B star, although younger than the A star, has
a much longer pulse period, about 2.77~s. Thirdly, the orbit
is viewed almost exactly edge-on, making the Shapiro delay easy to
observe \cite{ksm+06} and allowing eclipses of the A pulsar by the
magnetosphere of the B pulsar \cite{mll+04}. Recent observations of
the system have resulted in the detection of six relativistic effects,
including a measurement of the orbital decay that is an order of
magnitude more precise than the PSR B1913+16 determination and fully
consistent with the prediction from general relativity
(Kramer et al., in preparation). It never ceases to amaze me that,
despite the development of many alternative theories of relativistic
gravity since general relativity, Einstein seems to have got it right
the first time.

\section{Future prospects}
The past 50 years of pulsar astronomy have been more than interesting but
there is much to look forward to. Large new radio telescope such as
the recently commissioned Five hundred metre Aperture Spherical
Telescope (FAST) in Guizhou province, China \cite{nlj+11} will enable
the discovery and study of many more pulsars, as well as increasing
the precision of observations of currently known pulsars. Further in
the future, the Square Kilometre Array \cite{cr04} will provide
unrivalled sensitivity and resolution for essentially all radio
astronomy applications. New X-ray telescopes such as eRosita, to be
launched next year \cite{pab+16}, and Athena, due for launch in 2028
\cite{abf+15}, will provide wide-field spectroscopic imaging with
unprecedented sensitivity and resolution. Existing radio telescopes will be
enhanced with new wideband receivers and more sophisticated
signal-processing systems. All of these and more will surely probe deeper
into the secrets of pulsars, uncovering many things that are not even
dreamt of yet. The future of pulsar astronomy and astrophysics is bright!

{\Large
  \centerline{\foreignlanguage{russian}{\bf Большое спасибо!}}
}

\ack It has been a privilege and a pleasure to have travelled through
50 years of pulsar astronomy and astrophysics and I thank all those
colleagues and friends who have helped me along the way.

\section*{References}

%\bibliographystyle{iopart-num}
%\bibliography{journals,modrefs,psrrefs,crossrefs}

\providecommand{\newblock}{}

\end{document}